\documentclass[aps,prl,reprint,superscriptaddress]{revtex4-2}
\usepackage{amsmath}
\usepackage{graphicx}
\usepackage{bbm}

\usepackage{graphicx}
\usepackage{dcolumn}
\usepackage{bm}

\makeatletter

\begin{document}
\title{Hybrid photon blockade with hyperradiance in two-qubit cavity QED system}

\author{Zhuorui Wang}
\email[]{tungstenzr@outlook.com}
\affiliation{School of Microelectronics and Data Science, Anhui University of Technology, Ma'anshan, Anhui 243032, China}

\author{Jun Li}
\email[]{jli@ahut.edu.cn}
\affiliation{School of Microelectronics and Data Science, Anhui University of Technology, Ma'anshan, Anhui 243032, China}

\date{\today}

\begin{abstract}
	We investigate a hybrid photon blockade (HPB) scheme in a driven two-qubit cavity QED system arising from the combination of eigenenergy-level anharmonicity (ELA) and quantum destructive interference (QDI). By tuning the detuning of a single qubit and pumping field, we identify precise parametric regimes that fully integrate the advantages of high brightness in ELA-based conventional photon blockade and strong antibunching in QDI-based unconventional photon blockade. Interestingly, these regimes are accompanied by hyperradiance, indicating that inter-emitter correlations give rise to enhanced collective emission. The HPB mechanism exhibits parametric generality across varying coupling asymmetries and remains accessible via detuning control, offering a feasible route for generating high-quality single-photon source in diverse quantum platforms.
\end{abstract}

\maketitle

	Photon blockade (PB) is a paradigmatic nonlinear optical effect \cite{birnbaum2005photon} that underlies deterministic single-photon generation \cite{liew2010single, faraon2008coherent} required for quantum key distribution, repeater architectures, and photonic quantum computing \cite{castelletto2008heralded, li2016demand, chen2022photon, couteau2023applications}. PB has been extensively investigated across various platforms, including cavity quantum electrodynamics (QED) \cite{ hennrich2005transition, hamsen2017two}, cavity optomechanics \cite{rabl2011photon, nunnenkamp2011single}, circuit QED \cite{hoffman2011dispersive, blais2021circuit}, quantum dot QED \cite{tang2015quantum, snijders2018observation}, and waveguide QED \cite{lu2025chiral}. To date, a variety of studies have embedded PB into systems such as nonreciprocity in spinning resonators induced by the Fizeau light-dragging effect \cite{huang2018nonreciprocal, liu2023nonreciprocal, wang2023squeezing, li2019nonreciprocal}, non-Hermitian systems with exceptional points \cite{huang2022exceptional}, topological edge states \cite{li2024enhancement, zhang2025distinct} and Friedrich-Wintgen bound states in the continuum \cite{deng2025enhancement}, while the blockade mechanisms have been extended to other bosonic platforms, including phonon blockade and magnon blockade \cite{liu2010qubit, zhao2020simultaneous, falch2025magnon}.
	
	Various physical mechanisms have been proposed to realize PB \cite{zhou2025universal, lin2020kerr}, among which conventional photon blockade (CPB) and unconventional photon blockade (UPB) constitute two fundamental pathways \cite{flayac2017unconventional, hou2019interfering}. CPB arises from the eigenenergy-level anharmonicity (ELA) in strongly nonlinearity systems \cite{birnbaum2005photon}, whereas UPB relies on quantum destructive interference (QDI) between distinct excitation pathways, which can occur even in the weak nonlinear regime \cite{liew2010single, bamba2011origin, flayac2013input}. However, single-mechanism sources often face a fundamental trade-off between purity and brightness. To overcome this limitation, recent research has pivoted toward hybrid photon blockade (HPB) schemes that combine ELA-based mechanism and QDI-based mechanism interference within a unified platform \cite{liang2020photon, zhu2021hybrid, zhu2023strong, qiao2024strongly}. One approach employs a two-qubit cavity QED system with dipole–dipole interaction, where ELA and QDI act cooperatively to achieve HPB \cite{zhu2021hybrid}. Another method uses a hybrid two-cavity system with Kerr nonlinearity, in which CPB and UPB coexist under suitable parameter regimes \cite{zhu2023strong}. A universal photon blockade scheme has been proposed using a two-photon Jaynes-Cummings model, in which both conventional and unconventional mechanisms produce photon antibunching across a broad nonlinear parameter range \cite{zhou2025universal}. These developments collectively highlight the potential of hybrid schemes to overcome limitations inherent to single mechanism approaches.
	
	Beyond single-emitter systems, the interplay between PB and collective emission, such as superradiance and hyperradiance, facilitates the enhancement of quantum correlations through multi-emitter interactions \cite{pleinert2017hyperradiance,li2021electromagnetic, li2022squeezed}. In this Letter, we investigate a HPB scheme in a driven two-qubit cavity QED system \cite{zhu2017collective, zhu2021hybrid, hou2019interfering, zhao2019multiphoton, radulaski2017photon, reimann2015cavity} that exploits the interplay for ELA and QDI. By tuning the frequency of one qubit and its detuning relative to the cavity mode, we identify the precise parametric intersection at which the two mechanisms cooperatively reinforce each other, thereby enabling a simultaneous high-quality of purity and brightness at the optimal HPB point. Through an evaluation of the radiance witness $R$ across different regimes, we demonstrate that this high-performance blockade is accompanied by hyperradiant collective behavior  \cite{pleinert2017hyperradiance, li2021electromagnetic}, indicating that inter-emitter correlations drive the system into an intensified emission regime. Furthermore, we verify the parametric generality of the HPB mechanism across a wide range of coupling strength ratios between the two qubits, which remains accessible via independent control of the qubit detunings.

\begin{figure}[]
		\centering
		\includegraphics[width=\linewidth]{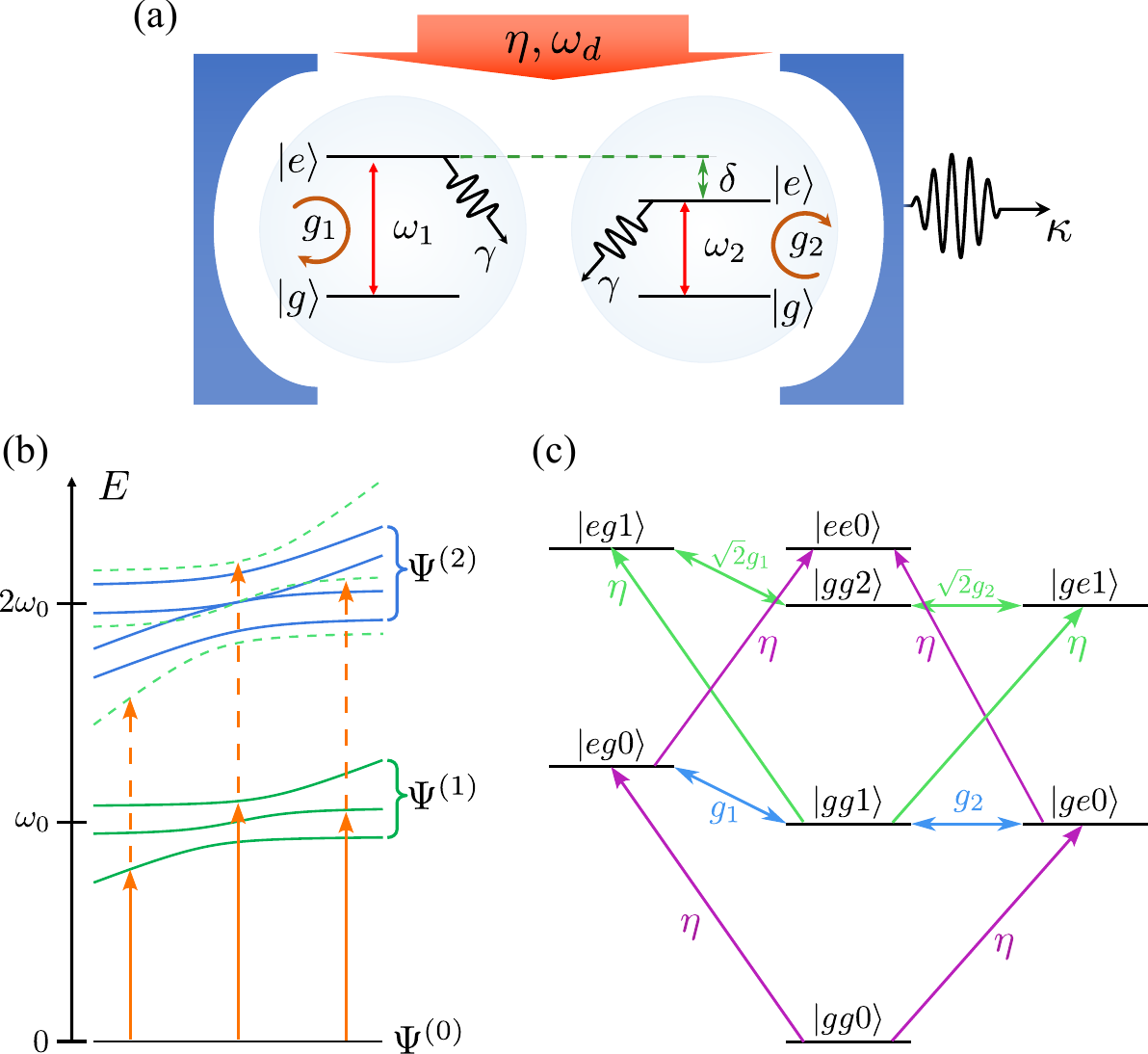}
		\caption{(a) Schematic diagram of the driven two-qubit cavity QED system. qubit 1 and qubit 2 are coupled to a single-mode cavity with coupling strengths $g_{1}$ and $g_{2}$ with the same spontaneous decay rate $\gamma$ , while the cavity decays at rate $\kappa$. The frequency detuning between the qubits is defined as $\delta$. (b) Dressed-state spectrum versus the qubit detuning $\delta$. Green (blue) solid curves denote the single- (two-) excitation subspaces. (c) Interfering transition pathways in the excitation process. The three colored arrows represent three sets of QDI channels.
		}
		\label{fig_1}
	\end{figure}

	We consider a driven two-qubit cavity QED system consisting of a single-mode cavity coherently coupled to two different two-level qubits, hereafter referred to as qubit 1 and qubit 2. As shown in Fig.\ref{fig_1}(a), the transition frequencies and cavity coupling strengths of the two qubits are denoted by $\omega_{1}, g_{1}$ and $\omega_{2}, g_{2}$, respectively. The frequency mismatch between the two qubits is characterized by $\delta=\omega_{1}-\omega_{2}$. Without loss of generality, qubit 2 is assumed to be resonant with the cavity mode, i.e. $\omega_{2}=\omega_{c}\equiv\omega_{0}$. Both qubits are coherently driven by an external field with frequency $\omega_{d}$ and driving strength $\eta$.
	
	In a frame rotating at the drive frequency $\omega_{d}$, the Hamiltonian of the system (setting $\hbar=1$) can be written as
	\begin{equation}
		\begin{aligned}
			H
			=& \left( - \Delta + \delta \right)\sigma_{1}^{+}\sigma_{1}^{-} - \Delta \sigma_{2}^{+}\sigma_{2}^{-}  - \Delta a^{\dagger}a  \\
			&+ \sum_{j=1}^{2} g_{j}\left( \sigma_{j}^{+}a + \sigma_{j}^{-}a^{\dagger} \right)
			+ \eta \sum_{j=1}^{2}\left( \sigma_{j}^{+} + \sigma_{j}^{-} \right),
		\end{aligned}
		\label{eq_Hamiltonian}
	\end{equation}
	where $a$ ($a^{\dagger}$) is the annihilation (creation) operator of the cavity mode, and $\sigma_{j}^{\pm}$ are the raising and lowering operators of qubit~$j$ $(j=1,2)$. Here the detuning parameter $\Delta\equiv\omega_{d}-\omega_{0}$ simultaneously describes the drive detuning for both qubit 2 and the cavity. Based on Born-Markov approximation, the dynamical behavior of this open qutantum system can be governed by the master equation
	\begin{equation}
		\dot{\rho} = -i\left[H, \rho\right] + \mathcal{L}[a] + \sum_{j=1}^{2}\mathcal{L}[\sigma_{j}^{-}],
		\label{eq_masterequation}
	\end{equation}
	where $\rho$ is the density matrix operator of this two-qubit cavity system, while the Liouvillian superoperators $\mathcal{L}[a] = 0.5\kappa \left( 2a\rho a^\dagger - a^\dagger a \rho - \rho a^\dagger a \right)$ and $\mathcal{L}[\sigma_j^-] = 0.5\gamma_j \left( 2\sigma_j^-\rho\sigma_j^+ - \sigma_j^+\sigma_j^-\rho - \rho\sigma_j^+\sigma_j^- \right)$ describes the cavity decay at rate $\kappa$ and spontaneous decay of the excited state of both qubits at rate $\gamma$, respectively.	
	By numerically solving Eq.\ref{eq_Hamiltonian} and \ref{eq_masterequation} for the density matrix $\rho$ using the QuTiP toolbox in Python \cite{johansson2012qutip}, we determine the equal-time second-order correlation function $g^{(2)}(0) = \langle a^\dagger a^\dagger a a \rangle / \langle a^\dagger a \rangle^2$, as well as the radiance witness $R$, to characterize the photon statistics and collective emission behavior.

	In the absence of driving and dissipation, we consider the dressed-state energy spectrum as a function of $\delta$ , as shown in Fig.\ref{fig_1}(b). In the single- and two-excitation subspaces, the eigenvalues are determined by the following equations
	\begin{equation} 
		(\varepsilon_{1} - \delta)\left(\varepsilon_{1}^{2} - g_{2}^{2}\right) - g_{1}^{2} \varepsilon_{1} = 0, 
		\label{eq_Eigenvalues1} 
	\end{equation}
	and
	\begin{equation} 
		\begin{aligned} 
			&\varepsilon_{2}^4 + \delta^2 (\varepsilon_{2}^2 - 2 g_2^2) + 2 (g_1^2 - g_2^2)^2 - 
			3 \varepsilon_{2}^2 (g_1^2 + g_2^2) \\
			&+ \delta \varepsilon_{2} (-2 \varepsilon_{2}^2 + 3 g_1^2 + 4 g_2^2) = 0, 
		\end{aligned} 
		\label{eq_Eigenvalues2} 
	\end{equation}
	respectively, where $\varepsilon_{j}=E_{j}-j\omega_{0}$ $(j=1,2)$, and $E_{j}$ denotes the eigenvalue in the $j$-excitation subspace. The resulting spectrum forms an anharmonic ladder-type structure. Across a broad range of $\delta$, a substantial energy mismatch exists between the doubled single-excitation eigenvalues $2E_1$ and the two-excitation eigenvalues $E_2$, particularly for the upper and lower dressed branches. Even the central dressed branch exhibits a pronounced mismatch away from the qubit-resonance condition. Such intrinsic ELA feature implies that when the pump field is resonant with a single-excitation state, the transition to the two-excitation subspace is  blocked, as indicated by the orange arrow in Fig.\ref{fig_1}(b).

	Complementary to ELA, the system also supports QDI-induced UPB. In the weak driving limit $(\eta \ll \kappa, \gamma)$, the population of higher excitation states is negligible, permitting the truncation of the Hilbert space to the two-excitation subspace. Using the Schrödinger equation, the probability amplitudes $C_{\alpha\beta n}$ (where $\alpha, \beta \in \{g,e\}$ denote the qubit states and $n$ represents the cavity photon number) evolve according to the following equations
	\begin{equation}
		\begin{aligned}
			&\mathrm{i}\dot{C}_{eg0} = \tilde{\Delta}_{q1} C_{eg0} + \eta C_{gg0} + \eta C_{ee0} + g_1 C_{gg1}, \\
			&\mathrm{i}\dot{C}_{ge0} = \tilde{\Delta}_{q2} C_{ge0} + \eta C_{gg0} + \eta C_{ee0} + g_2 C_{gg1}, \\
			&\mathrm{i}\dot{C}_{gg1} = \tilde{\Delta}_c C_{gg1} + \eta (C_{eg1} + C_{ge1}) + g_1 C_{eg0} + g_2 C_{ge0}, \\
			&\mathrm{i}\dot{C}_{ee0} = (\tilde{\Delta}_{q1} + \tilde{\Delta}_{q2}) C_{ee0} + \eta (C_{eg0} + C_{ge0}) + g_2 C_{eg1} + g_1 C_{ge1}, \\
			&\mathrm{i}\dot{C}_{eg1} = (\tilde{\Delta}_{q1} + \tilde{\Delta}_c) C_{eg1} + \sqrt{2}g_1 C_{gg2} + g_2 C_{ee0} + \eta C_{gg1}, \\
			&\mathrm{i}\dot{C}_{ge1} = (\tilde{\Delta}_{q2} + \tilde{\Delta}_c) C_{ge1} + \sqrt{2}g_2 C_{gg2} + g_1 C_{ee0} + \eta C_{gg1}, \\
			&\mathrm{i}\dot{C}_{gg2} = 2\tilde{\Delta}_c C_{gg2} + \sqrt{2}g_1 C_{eg1} + \sqrt{2}g_2 C_{ge1},
		\end{aligned}
		\label{eq_probability amplitudes}
	\end{equation}
	where $\tilde{\Delta}_{q1} = -(\Delta - \delta) - i\gamma/2$, $\tilde{\Delta}_{q2} = -\Delta - i\gamma/2$, and $\tilde{\Delta}_c = -\Delta - i\kappa/2$ are the complex detunings incorporating dissipation. Under the steady-state approximation and assuming ${\Delta, \delta \gg \kappa, \gamma},{C_{gg0} \approx 1}$, $C_{gg2}$ in symmetric-coupling case ($g_1=g_2$) is given by 
	\begin{equation}
		C_{gg2} \approx \frac{- \sqrt2 \eta^2  g_1^2\left(\delta-2\Delta\right)\left(\delta-3\Delta\right)\left(\delta-4\Delta\right)}{2F}
		\label{eq_C_gg2}
	\end{equation}
	where $F = (8\Delta^4-8\delta\Delta^3+2\delta^2\Delta^2-12g_1^2\Delta^2+7\delta g_1^2\Delta-\delta^2g_1^2) \times
	\left(\Delta^3-\delta\Delta^2-2g_1^2\Delta+\delta g_1^2\right)$. Given that the second-order correlation function reduces to $g^{(2)}(0) \approx 2|C_{gg2}|^2/|C_{gg1}|^4$ in this limit, ideal antibunching requires $C_{gg2} \rightarrow 0$. This condition yields three linear region: $\delta = 2\Delta$, $\delta = 3\Delta$, and $\delta = 4\Delta$. These distinct quantum destructive interference channels are indicated by the colored arrows in Fig.\ref{fig_1}(c).
	
	\begin{figure}[]
		\centering
		\includegraphics[width=\linewidth]{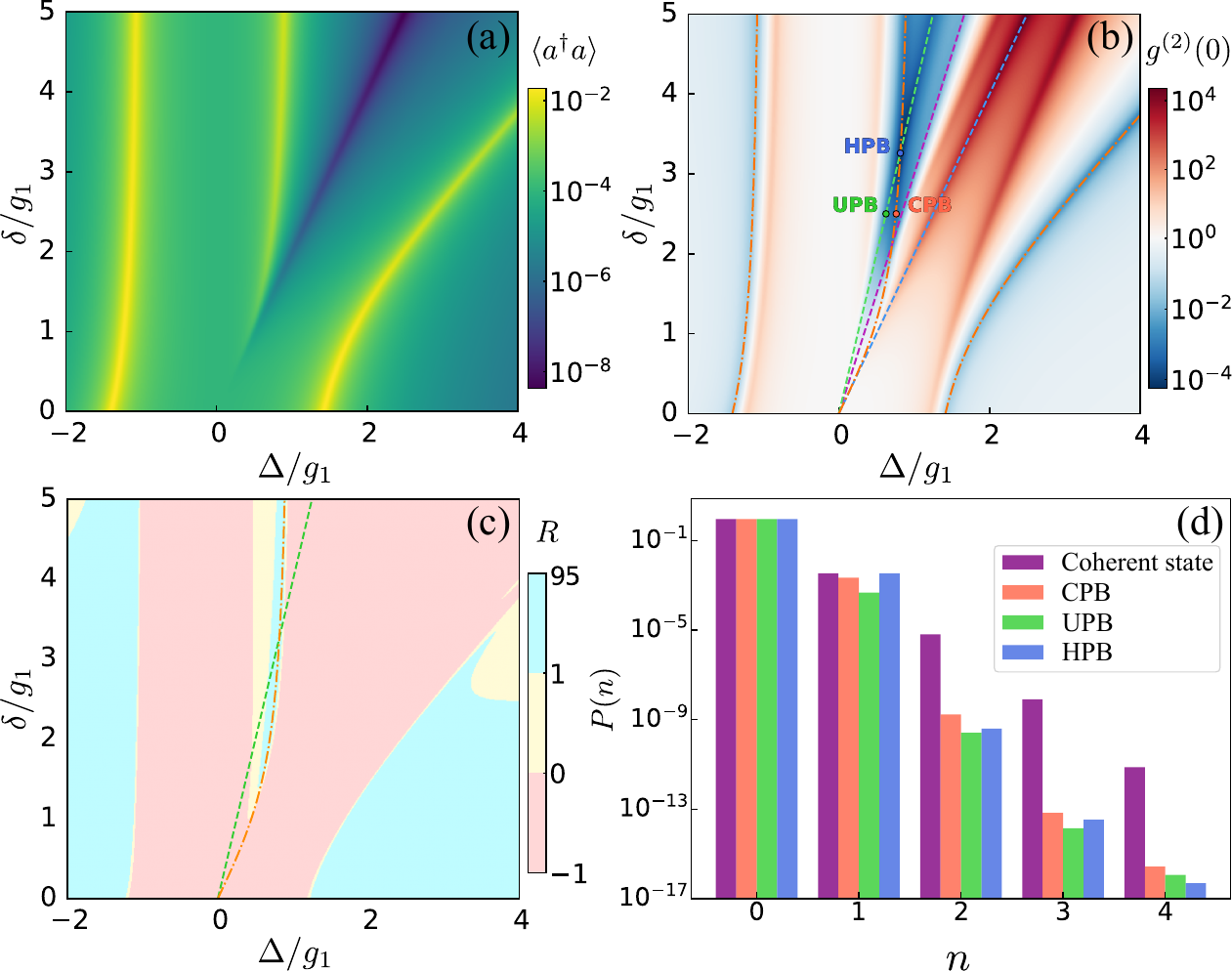}
		\caption{(a) Mean photon number $\langle a^\dagger a \rangle$, (b) second-order correlation function $g^{(2)}(0)$, and (c) radiance witness $R$ as functions of the normalized detunings $\Delta/g_{1}$ and $\delta/g_{1}$. The dashed lines denote the three QDI channels identified in Fig.\ref{fig_1}(a), while The dot-dashed lines delineate the blockade regions governed by the ELA mechanism. (d) Photon number distribution $P(n)$ in the Fock state for the three blockade regimes highlighted in (b). Here, we set the parameters to $g_{1}=g_{2}=10\kappa$, and $\eta/\kappa=0.1$.
		}
		\label{fig_2}
	\end{figure}
	
	To elucidate the roles of ELA and QDI more clearly, Fig.\ref{fig_2}(a)(b) presents $\langle a^\dagger a\rangle$ and $g^{(2)}(0)$ versus the normalized detuning $\Delta/g_1$ and $\delta/g_1$ under the symmetric-coupling condition $g_1=g_2$. As shown in Fig.\ref{fig_2}(a), $\langle a^\dagger a \rangle$ exhibits three distinct branches of local maxima, where the mean photon number is enhanced by 1 to 2 orders of magnitude compared to surrounding regions, each corresponding to the resonant excitation of the three single-excitation eigenstates, i.e., $\omega_d=E_1$. Within the region where $\delta \approx 0$, the middle branch corresponds to a dark state that decouples to other states of the system; as $\delta$ increases, this dark state gradually transitions into a bright state. Besides these ELA-based branches, Fig.\ref{fig_2}(a) also reveals a branch with minimum in $\langle a^\dagger a\rangle$ along $\delta=2\Delta$, which corresponds to one of the QDI conditions identified in Fig.\ref{fig_1}(c). This minimum arises from destructive interference between the two excitation pathways $|gg0\rangle\overset{\eta}{\longrightarrow}|eg0\rangle \overset{g_1}{\longleftrightarrow}|gg1\rangle$ and $|gg0\rangle\overset{\eta}{\longrightarrow}|ge0\rangle \overset{g_2}{\longleftrightarrow}|gg1\rangle$, leading to blockade of the single-photon excitation. Consistently, Fig.\ref{fig_2}(b) shows that this line does not yield antibunching, since $g^{(2)}(0)>1$. In contrast, the QDI conditions $\delta=3\Delta$ and $\delta=4\Delta$ coincide with minimum of $g^{(2)}(0)$ and thus correspond to UPB. The latter originates from interference between the two two-photon excitation pathways from $|gg1\rangle$ to $|gg2\rangle$, whereas the former involves the intermediate state $|ee0\rangle$. 
	
	Crucially, the UPB region intersect with the ELA branches at varied $\delta$ values, giving rise to hybrid PB phenomenon. At the primary HPB point where $\delta = 4\Delta$, $g^{(2)}(0)$ reaches $10^{-4}$, a reduction of approximately 1.5 orders of magnitude compared to isolated UPB points. Simultaneously, the mean photon number remains near $0.0035$, preserving the high-brightness advantage of CPB and demonstrating a strong PB effect. Meanwhile, Fig.\ref{fig_2}(d) compares the cavity photon-number distribution $P(n)$ for the coherent state, CPB, UPB, and HPB points marked in \ref{fig_2}(b). The HPB scheme achieves a $P(1)$ value comparable to that of a coherent state and an order of magnitude higher than that of UPB. Meanwhile, the $P(2)$ for HPB is minimized to a level similar to UPB, which is 4 orders of magnitude lower than the $P(2)$ of a coherent state. This photon-number distribution highlights the capability of HPB to simultaneously suppress multiphoton states and maintain a high single-photon occupancy.
	
	To further characterize the collective radiance behavior in the HPB regime, we introduce and calculate the radiance witness $R$ \cite{pleinert2017hyperradiance, zhao2019multiphoton}, defined as
	\begin{equation}
		R=\frac{\left\langle a^{\dagger} a\right\rangle_{2}-\sum_{i=1}^{2}\left\langle a^{\dagger} a\right\rangle_{1, i}}{\sum_{i=1}^{2}\left\langle a^{\dagger} a\right\rangle_{1, i}},
		\label{eq_R}
	\end{equation}
	where $\langle a^\dagger a \rangle_2$ and $\langle a^\dagger a \rangle_{1,i}$ quantify the mean photon numbers for the two-qubit and $i$th single-qubit in the cavity, respectively. $R < 0$ characterizes subradiance arising from radiative suppression, while $R > 1$ denotes hyperradiance which means that the collective emission from the coupled emitters is higher than that of a corresponding system of uncorrelated atoms. Due to the resonant enhancement of single-photon excitation accompanied by the suppression of higher-order processes, hyperradiance is accompanied with antibunching, which is also confirmed in our system. As shown in Fig.~\ref{fig_2}(c), the radiance witness $R$ reaches a value of about $R\approx 3$, indicating a pronounced collective emission enhancement. These results reveal that by tuning the qubit and drive tunings to the HPB points, one can achieve PB phenomenon that combine the benefits of high purity and brightness.
	
	\begin{figure}[]
		\centering
		\includegraphics[width=\linewidth]{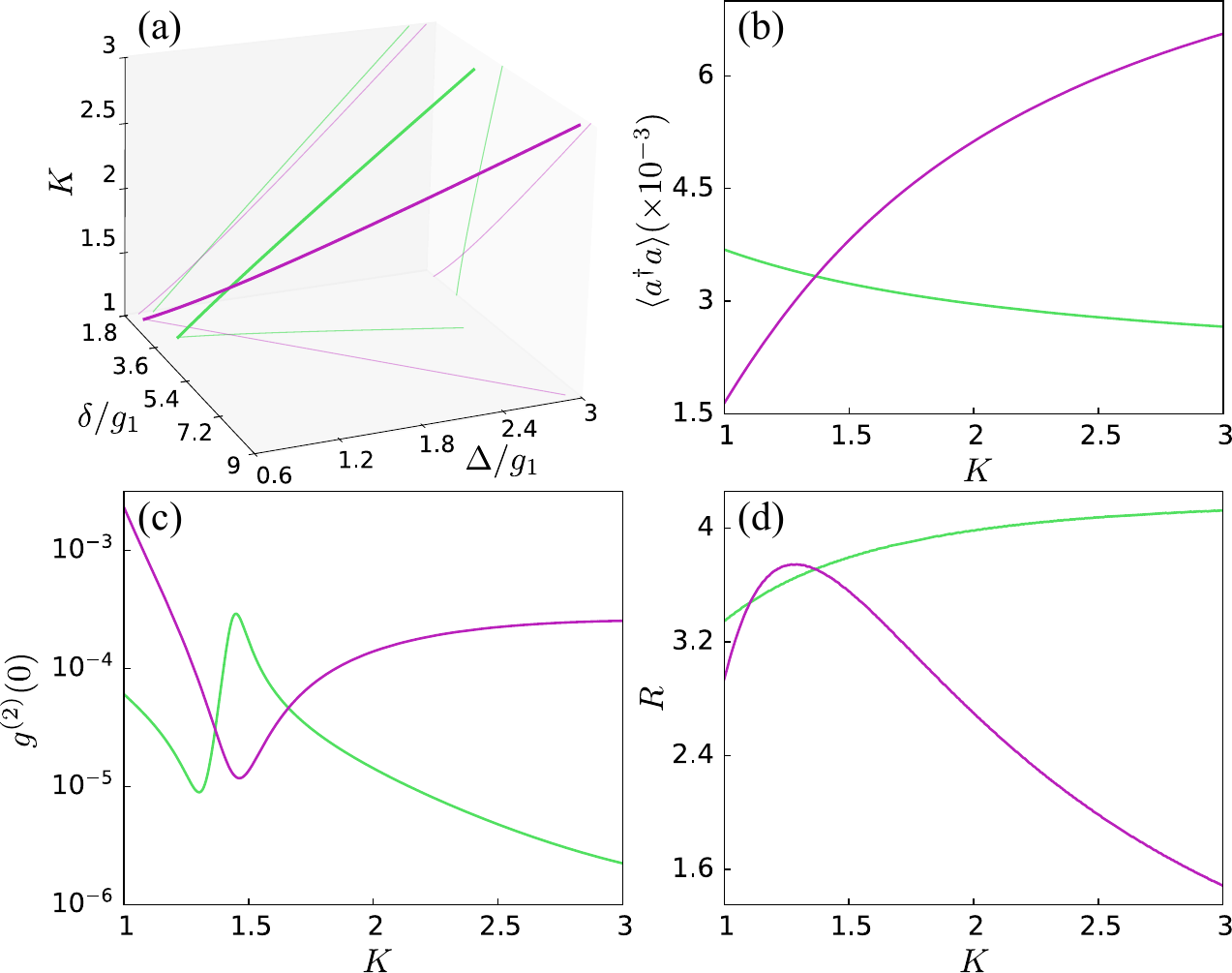}
		\caption{Characteristics of HPB under varying coupling asymmetries. (a) Three-dimensional trajectories of the HPB points in the $(\Delta/g_1, \delta/g_1, g_{2}/g_{1})$ parameter space. (b) mean photon number $\langle a^\dagger a \rangle$, (c) equal-time second-order correlation function $g^{(2)}(0)$, and (d) radiance witness $R$. The green and purple solid lines correspond to the primary and secondary HPB series, respectively, which are colored to match the two QDI channels identified in Fig.\ref{fig_1}(c). Here, we set $K = g_{2}/g_{1}$. All other system parameters remain identical to those used in Fig.\ref{fig_2}.
		}
		\label{fig_3}
	\end{figure}
	
	We examine the parametric generality of the proposed scheme by investigating the evolution of HPB under varying coupling asymmetries in Fig.\ref{fig_3}. As illustrated in Fig.\ref{fig_3}(a), increasing the coupling ratio $K = g_{2}/g_{1}$ from 1 to 3 causes both HPB series to migrate monotonically toward larger detuning regimes in the parameter space. This persistent migration confirms that the hybrid blockade is not a singular parameter occurrence but a trackable phenomenon that can be dynamically preserved via independent frequency control. Notably, we derive the analytical trajectories of these HPB points as functions of the coupling ratio. The primary series (green) follows the trajectory defined by $\Delta/g_{1} = \sqrt{2K^{3}/(2K + 1)}$ and $\delta = \Delta(1 + 1/K)^2$, while the secondary series (purple) adheres to $\Delta/g_{1} = \sqrt{K^{2} - 1/2}$ and $\delta = 3\Delta$. The latter strictly linear locking implies a fundamental invariance of the QDI channel relative to the system eigenenergies despite significant coupling mismatch.
	
	The blockade performance and its relation to collective emission are further characterized in Figs.\ref{fig_3}(b)–-\ref{fig_3}(d). In terms of photon purity, Fig.\ref{fig_3}(c) shows a competitive evolution between the two branches. The secondary branch achieves slightly lower $g^{(2)}(0)$ near $K \approx 1.5$, whuile the primary branch reaches its minimum at large $K$ as the coupling asymmetry increases. This behavior correlates with the evolution of collective emission. As shown in Fig.\ref{fig_3}(d), the radiance witness $R$ associated with the primary branch increases steadily with $K$, indicating a continuous enhancement of hyperradiant emission. By contrast, the secondary branch shows a nonmonotonic dependence, with $R$ reaching a maximum and then decreasing, suggesting a weakening of inter-emitter correlations. Meanwhile, Fig.\ref{fig_3}(b) shows that the mean photon number $\langle a^\dagger a \rangle$ of the primary branch decreases only gradually. Overall, these results demonstrate that HPB persists over a broad parameter range, with the primary branch offering low  and enhanced collective emission.
	
	In conclusion, we have shown and theoretically analyzed a HPB scheme in a driven two-qubit cavity QED system arising from the combination of ELA-baed CPB and QDI-based UPB. The required operating conditions are realized through the independent tuning of the qubit frequency and detuning, both of which are experimentally accessible parameters \cite{muller2015coherent, hamsen2017two, trivedi2019photon}. Numerical evaluations show that, within the optimal parameter regime, the HPB enables a strong suppression of $g^{(2)}(0)$ while maintaining a large mean photon number compared to both CPB and UPB. We further observe that this high-performance regime is accompanied by hyperradiant behavior, suggesting the presence of nontrivial inter-emitter correlations. The persistence of the hybrid blockade across varying coupling ratios confirms its parametric generality. The results provide a feasible framework for balancing photon statistics and emission efficiency in diverse cavity QED platforms.
	
	This work was supported by the Talent Project Scientific Research Foundation for High-level Talents of Anhui University of Technology (QD202425, QD202305).
	
\bibliography{ref}

\end{document}